\documentclass[aps,prl,twocolumn,showpacs,nofootinbib,superscriptaddress]{revtex4-1}
\usepackage[a4paper,top=1in,bottom=1in,left=0.5in,right=0.5in]{geometry}
\usepackage{amsmath,latexsym}
\usepackage{amssymb}
\usepackage{graphicx}
\usepackage{dcolumn}
\usepackage{bm}
\usepackage{xcolor}
\usepackage[%
  colorlinks=true,
  urlcolor=blue,
  linkcolor=blue,
  citecolor=blue
]{hyperref}
\usepackage{mathptmx}
\usepackage{tabularx}
\usepackage{etoolbox}

\newcommand{\beq}{\begin{eqnarray}}
\newcommand{\eeq}{\end{eqnarray}}
\def\point#1#2{{\tt #1}_{\mbox{\footnotesize #2}}}
\def\orbital{{\tt SO(3)_{\text{L}}}}
\def\spin{{\tt SO(3)_{\text{S}}}}
\def\gauge{{\tt U(1)_{\text{N}}}} 
\def\parity{{\tt P}} 
\def\time{{\tt T}}

\def\He{{$^3$He}}

\def\Heb{{$^3$He-B}}
\makeatletter
\let\cat@comma@active\@empty
\makeatother
\begin{document}

\title{
Bosonic Surface States and Acoustic Spectroscopy of Confined Superfluid $^3$He-B
}

\author{Takeshi Mizushima}
\email{mizushima@mp.es.osaka-u.ac.jp}
\affiliation{Department of Materials Engineering Science, Osaka University, 
             Toyonaka, Osaka 560-8531, Japan}

\author{J. A. Sauls}
\email{sauls@northwestern.edu}
\affiliation{Department of Physics and Astronomy, Northwestern University, 
             Evanston, Illinois 60208}

\date{\today}

\begin{abstract}
Using an effective field theory  we study the low-lying bosonic excitations and their couplings to phonons at ultrasonic frequencies in superfluid $^3$He-B under strong confinement. We show that confinement induces a rich spectrum of low-lying bosons, including surface-bound bosonic states, as well as fine structure of long-lived massive bosons. 
Under sufficiently strong confinement we find a dynamical instability of the \Heb\ film: the frequency of the surface-bound boson softens at finite wavevector, then develops a pole in the upper half of the complex frequency plane, signalling a dynamical instability of the translationally invariant superfluid vacuum towards pair-density-wave ``crystallization''.
We discuss the signatures and observability of the low-lying bosonic spectrum based on analysis of the ultrasound attenuation from resonant excitation of the bosonic modes.
We also note that surface-bound bosonic modes are not unique to $^3$He-B, but are expected to be common to unconventional superconductors with a multi-component order parameter.

\end{abstract}


\maketitle

{\it Introduction.}--- The low-energy physics of DIII topological superconductors (SCs), is governed by two key ingredients -- bosons as collective modes of the ordered state, and helical Majorana fermions (MFs) as topologically protected Bogoliubov quasiparticles. The helical MFs emerge on the surface as a hallmark of nontrivial topology in momentum space~\cite{miz16,sat16,vol16}, while the spectrum of bosonic modes reflects the symmetry of the superfluid vacuum~\cite{sau00a}. Among them, Nambu-Goldstone (NG) and massive Higgs bosons are the centerpieces of dynamics and transport phenomena, as they involve the coherent motion of a macroscopic fraction of particles. 

\begin{figure}[b!]
\includegraphics[width=85mm]{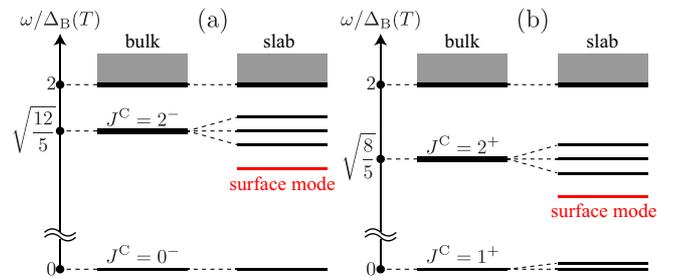}
\caption{Bosonic energy-level spectra for ${\rm C}=-$ (a) and ${\rm C}=+$ (b) in the bulk and for $^3$He-B confined in a slab. New surface bosonic excitations appear in confined geometry. In confined geometry some of $J^{\rm C}=1^+$ NG bosons acquire a small mass and become pseudo-NG bosons.}
\label{fig:levels}
\end{figure}

The B phase of superfluid $^3$He, a spin-triplet $p$-wave superfluid with an isotropic gap $\Delta$, has served as a prototype for studying transport phenomena mediated by a variety of low-lying bosons, as fermions are frozen out at low temperatures.  The B-phase of $^3$He spontaneously breaks relative spin-orbit rotation symmetry, while preserving joint rotation symmetry in spin and orbital spaces. Thus, the 18 bosonic modes of $^3$He-B are categorized in terms of the total angular momentum, $J \in \{0, 1,2\}$, and parity under particle-hole conversion, ${\rm C}=\pm$. The spectrum includes 4 NG bosons: the $J^{\rm C}=0^-$phase mode and 3 $J = 1^+$ spin-orbit modes, as well as 14 massive Higgs modes (c.f. Fig.~\ref{fig:levels}). 

The $J^{\rm C}=2^{\pm}$ quintets are long-lived Higgs bosons with masses smaller than the threshold energy, $2\Delta$, for decay into unbound fermion pairs. The $J^{\rm C}=2^-$ bosons couple strongly to mass current fluctuations~\cite{wol73,mak74,nag75,moo93}, while the coupling of $J^{\rm C}=2^+$ bosons to mass currents becomes active through the weak violation of the particle-hole symmetry by the normal fermionic vacuum~\cite{koc81}. Indeed, the masses of the $J^{\rm C}=2^{\pm}$ modes have been precisely measured using resonant longitudinal ultrasound absorption spectroscopy~\cite{gia80,mas80,ave80,sau17}, transverse ultrasound velocity spectroscopy, and acoustic Faraday rotation of transverse sound measurements~\cite{lee99,dav06,dav08,dav08c,col13}. In addition to the $J=2$ bosons, the $J^{\rm C}=1^{+}$ NG/pseudo-NG bosons were detected through the decay of a Bose-Einstein condensate of magnons~\cite{zav16}.

 Superfluid $^3$He-B exhibits {\emph{emergent} DIII topology as a consequence of spontaneous symmetry breaking of parity and separate spin, orbital and gauge symmetries by the fermionic vacuum to the sub-group of combined spin and orbital rotations and time-reversal, i.e. $\spin\times\orbital\times\gauge\times\time\times\parity\rightarrow\point{SO(3)}{J}\times\time$. A consequence of the bulk-edge correspondence is the spectrum of massless Majorana fermions confined on the surface of $^3$He-B. The signatures and observable consequences of the surface Majorana spectrum are a subject of current theoretical and experimental investigation~}\cite{miz16,oku12,miz15}. In this Letter we show that the broken symmetries exhibited by $^3$He-B, combined with dimensional confinement, also lead to a new spectrum of surface-bound bosonic modes. (see Fig.~\ref{fig:levels}).

We develop an effective field theory that incorporates the confinement potential, and captures all the features of the space-time fluctuations of the bosonic fields.  We focus on two key consequences of confinement: (i) surface depairing and (ii) additional symmetry breaking by the confining potential. The former leads to distortion of the equilibrium order parameter near the surface. We demonstrate that such local distortions generate attractive potentials for the bosonic fields. As a result surface-bound bosons emerge. The existence of surface-bound bosons is not unique to $^3$He-B, but is expected to be common to unconventional SCs with a multi-component order parameter, as well as inhomogeneous SCs hosting topological defects \cite{ham16}. We also show that one of the surface modes softens as the confinement approaches a critical thickness, signaling a dynamical instability of translationally invariant $^3$He-B within the slab to a crystalline-ordered superfluid \cite{vor07}. Symmetry reduction by the confinement potential also partially lifts the fivefold degeneracy of the bulk $J = 2$ Higgs modes, similar to crystal-field splitting of electronic levels of ions confined in the solid state.

{\it Effective field theory.}--- An effective field theory for superfluid $^3$He-B is based on the underlying fermionic spectrum of topological superfluid $^3$He-B, i.e. that low-lying quasiparticle states are decomposed into surface-bound helical MFs for energies $|E| < \Delta$, and extended states with $|E| >\Delta$. The sub-gap part of the spectrum generates the action for the Majorana fields, while the bosonic spectrum develops from the negative energy continuum by the formation of bound-fermions from the $p$-wave, spin-triplet interaction in the Cooper channel \cite{sau00a}. The pair condensate and bosonic excitations are described by the order parameter for spin-triplet $p$-wave superfluids, which is represented by the tensor field, $\mathcal{A}_{\mu i}$, with spin index $\mu$ and momentum index $i$ ($\mu, i=x,y,z$). The action for the bosonic fields is obtained from the bosonization of the fermionic action and the expansion of the bosonized action with respect to small fluctuations around the equilibrium order parameter, $A_{\mu i}$, that is, $|\mathcal{D}_{\mu i}| \equiv |\mathcal{A}_{\mu i}-A_{\mu i}| \ll \Delta $. The resulting effective action reduces to a {\it Lorentz invariant} form \cite{miz18},
\beq\label{eq:L}
S_{\rm B}=\int d^4x 
          \left\{\tau\,{\rm Tr}\left[\partial_t\mathcal{A}^{\ast}\partial_t\mathcal{A}\right]
         -       f_{\rm GL}[\mathcal{A}^{\ast},\mathcal{A}]
	  \right\}
\,,
\eeq
which is an extension of the Higgs model~\cite{hig64} to tensor fields. The effective inertia for dyamical fluctuations of $\mathcal{A}_{\mu i}$ is microscopically obtained as $\tau=3\beta_0=\frac{7\zeta(3)N_{\rm F}}{40[\pi k_{\rm B}T_{\rm{c}0}]^2}$, in the weak-coupling BCS approximation~\cite{miz18}, where $N_{\rm F}$ is the density of states of the normal fermionic vacuum and $T_{\rm{c}0}$ is the bulk superfluid transition temperature. 
For $|A_{\mu i}|\ll k_{\rm B}T_{\rm{c}_0}$, the potential term reduces to the Ginzburg-Landau (GL) functional, 
${f}_{\rm GL}=  f_{\rm b} + f_{\rm g}$, where the
bulk and gradient free energy terms are given in the weak-coupling approximation by
$f_{\rm b} =
-\frac{{N}_{\rm F}}{3}(1-\frac{T}{T_{{\rm c}0}}){\rm Tr}\!(\mathcal{A}\mathcal{A}^{\dag}) 
- \frac{3}{5}\beta_0 | {\rm Tr}\!(\mathcal{A}\mathcal{A}^{\rm T})|^2
+ \frac{6}{5}\beta_0 [ {\rm Tr}\!(\mathcal{A}\mathcal{A}^{\dag})]^2 
+ \frac{6}{5}\beta_0 {\rm Tr}\![\mathcal{A}\mathcal{A}^{\rm T}(\mathcal{A}\mathcal{A}^{\rm T})^{\ast}] 
+ \frac{6}{5}\beta_0 {\rm Tr}\![ (\mathcal{A}\mathcal{A}^{\dag})^2] 
- \frac{6}{5}\beta_0 {\rm Tr}\![ \mathcal{A}\mathcal{A}^{\dag}(\mathcal{A}\mathcal{A}^{\dag})^{\ast}]$ 
and
$
f_{\rm g} = \frac{1}{5}N_{\rm F}\xi^2_0( \partial _i \mathcal{A}^{\ast}_{\mu j} \partial _i \mathcal{A}_{\mu j} 
+ \partial _i \mathcal{A}^{\ast}_{\mu j} \partial _j \mathcal{A}_{\mu i} 
+ \partial _i \mathcal{A}^{\ast}_{\mu i} \partial _j \mathcal{A}_{\mu j})
$, respectively. NB: $a^{\rm T}$ denotes the transpose of $a$, and $\xi_0=\hbar v_f/2\pi k_{\rm B} T_{c0}$ is the coherence length which varies from $\xi_0 \simeq 80-20\,\mbox{nm}$ over the pressure range, $p=0-34\,\mbox{bar}$.

The order parameter of bulk $^3$He-B is given by $A_{\mu i}=\Delta(T)\delta_{\mu i}$, which is scalar under $\point{SO(3)}{J}$. The equations of motion for $\mathcal{D}_{\mu i}$ obtained from the principle of least-action applied to Eq.~\eqref{eq:L} describe all bosonic modes in bulk $^3$He-B~\cite{sau17,miz18}.
Each $J$ sector of the bosonic spectrum satisfies Nambu's fermion-boson mass relation, $\sum_{\rm C}(M^{\rm C}_J)^2=(2\Delta)^2$, where $\Delta$ is the Bogoliubov fermion mass (gap)~\cite{nam85,vol13,sau17}. As shown in Fig.~\ref{fig:levels}, the ${\rm C}=-$ sector contains the NG mode for $J^{\rm C}=0^-$ and the massive $J^{\rm C}=1^-$ and $J^{\rm C}=2^-$ modes with $M^{-}_{1}=2\Delta$ and $M^{-}_2=\sqrt{12/5}\Delta$. Their charge conjugation partners with ${\rm C}=+$ include the $J^{\rm C}=0^{+}$ Higgs boson, the $J^{\rm C}=1^{+}$ NG bosons, and the $J^{\rm C}=2^{+}$ quintet with mass, $M_2^+=\sqrt{8/5}\Delta$.

\begin{figure}[t!]
\includegraphics[width=85mm]{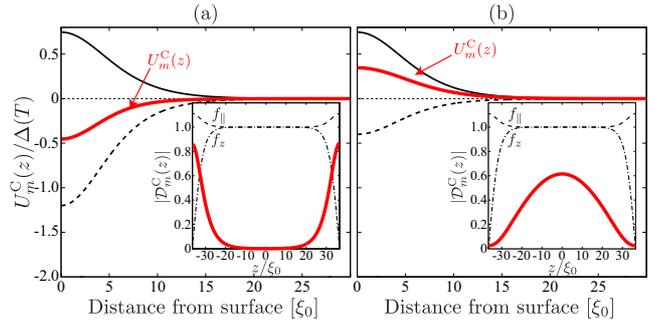}
\caption{Effective potentials, $U^{\rm C}_{m}(z)$, for $m=\pm 2$: (a) $(m,{\rm C})=(\pm 2,-1)$ and (b) $(\pm 2, +1)$, where the solid and broken lines (black) show the potenial constructed from $f_{\parallel}$ and $f_{z}$, respectively. The insets show the spatial profiles of the lowest excitation modes, $|\mathcal{D}^{\rm C}_{m,n=1}(z)|$, where the dashed and dashed-dotted curves correspond to the equilibrium order parameters. 
}
\label{fig:surface}
\end{figure}

{\it Surface-bound bosonic modes.}--- Consider superfluid $^3$He-B 
confined between two parallel boundaries with thickness $D$, where the $\hat{\bm z}$-axis is normal to the boundaries. The equilibrium order parameter is then given by
\beq\label{eq:bw}
A_{\mu i}(z) = \Delta (T) \left(
\begin{array}{ccc}
f_{\parallel}(z) & 0 & 0 \\ 0 & f_{\parallel}(z) & 0 \\ 0 & 0 & f_z(z)
\end{array}
\right)_{\mu i}
\,.
\eeq
The variables $f_{\parallel}$ and $f_z$ describe the the distortion of the order parameter by the confining potential. The spatial profiles of the order parameter for $D/\xi _0=73$ and $T/T_{{\rm c}0}=0.8$ are shown in the insets of Fig.~\ref{fig:surface}. We impose specular boundary conditions, $\partial_z f_{\parallel}=0$ and $f_z=0$ at the 
surfaces, $z=\pm D/2$.

For bosonic modes propagating with momentum $Q_{\parallel}$ parallel to the surface of the confining potential the equations of motions for $\mathcal{D}^{\rm C}_{\mu i}(z)e^{-i\omega t+iQ_{\parallel}x}$ are 
\beq\label{eq:H}
\mathcal{H}^{\rm C}_{\mu i,\nu j}(z,Q_{\parallel})\mathcal{D}^{\rm C}_{\nu j} (z,Q_{\parallel})
= 
\omega^2\mathcal{D}^{\rm C}_{\mu i} (z,Q_{\parallel})
\,,
\eeq
where $\mathcal{D}^{\rm C}_{\mu i}=\mathcal{D}_{\mu i}+{\rm C}\mathcal{D}^{\ast}_{\mu i}$ is odd (${\rm C}=-1$) and even (${\rm C}=+1$) under charge conjugation. The matrix elements are defined by the functional derivatives of $f_{\rm GL}$, $\mathcal{H}^{\rm C}_{\mu i,\nu j}=[\delta^2f_{\rm GL}/\delta\mathcal{D}^{\rm C}_{\mu i}\delta\mathcal{D}^{\rm C}_{\nu j}]_0$, evaluated in equilibrium.

The equilibrium state in Eq.~\eqref{eq:bw} is invariant under combined rotation in spin and orbital spaces about the $\hat{\bm z}$ axis, i.e. $\point{SO(2)}{J$_z$}$. Thus, Eq.~\eqref{eq:H} is block-diagonalized in terms of the eigenvectors of $\hat{J}_z$, with eigenvalues, $m_J=\{0,\pm 1,\pm 2\}$. In addition, Eq.~\eqref{eq:bw} is invariant under, $P_z$, inversion along $\hat{\bm z}$. Let $P = P_z\,R[\pi\hat{\bm z}]$ be the inversion operator combined with a $\pi$ rotation. The bosonic modes are then the eigenstates of the operator, $P\mathcal{D}_{\mu i}(z)=\mathcal{D}_{\mu i}(-z)=\pm \mathcal{D}_{\mu i}(z)$. The $m_J=0$ sector is further subdivided by parity under the mirror reflection symmetry combined with a $\pi$ rotation, $\{0^{+},0^-\}$, where the mirror plane contains the $\hat{\bm z}$ axis and is perpendicular to the surface. The full set of quantum numbers for the bosonic modes is given by $(m,P,{\rm C})=(\{0^+,0^-,\pm 1, \pm 2\},\pm,\pm)$. 

For $J^{\rm C}=2^{\pm}$ the $m=\{-2,-1,0,+1,+2\}$ modes obtained from Eq.~\eqref{eq:H} have bound-state solutions in confined geometry. Here we focus on the $m=\pm 2$ sector. For ${Q}_{\parallel}=0$, $\mathcal{D}^{\rm C}_{m=\pm 2}(z)$ is governed by the Schr\"{o}dinger-type equation
\begin{align}\label{eq:eom}
\left[-\lambda^2 \partial^2_z + U^{\rm C}_m(z)\right]\mathcal{D}^{\rm C}_{m}(z)
=
\left[(\omega^{\rm C}_{m})^2-(M^{\rm C}_J)^2\right]\mathcal{D}^{\rm C}_{m}(z)
\,,
\end{align}
 where $\lambda^2 \equiv \frac{7\zeta(3)\xi^2_0}{2(1-T/T_{\rm c0})}$.
The potential is determined by the equilibrium order parameter,
$U^{\rm C}_{\pm 2}(z)=\frac{16}{5}[f^2_{\parallel}(z)-1]-\frac{2(2-{\rm C})}{5}[1-f^2_z(z)]$.

Figures \ref{fig:surface}(a) and \ref{fig:surface}(b) show $U^{\rm C}_m(z)$ for $(m,{\rm C})=(\pm 2,-1)$ and $(\pm 2,+1)$ at $D/\xi _0 = 73$ and $T/T_{{\rm c}0}=0.8$, respectively. The local enhancement (suppression) of the parallel (perpendicular) component, $f^2_{\parallel}>1$ ($f^2_z<1$), near the surface generates the repulsive (attractive) potential in the case of specular surfaces. However, the total potential becomes attractive, and thus a bound state solution with excitation energy $\omega< M^-_2=\Delta\sqrt{12/5}$ exists only in the ${\rm C}=-$ sector.

\begin{figure}[t!]
\includegraphics[width=85mm]{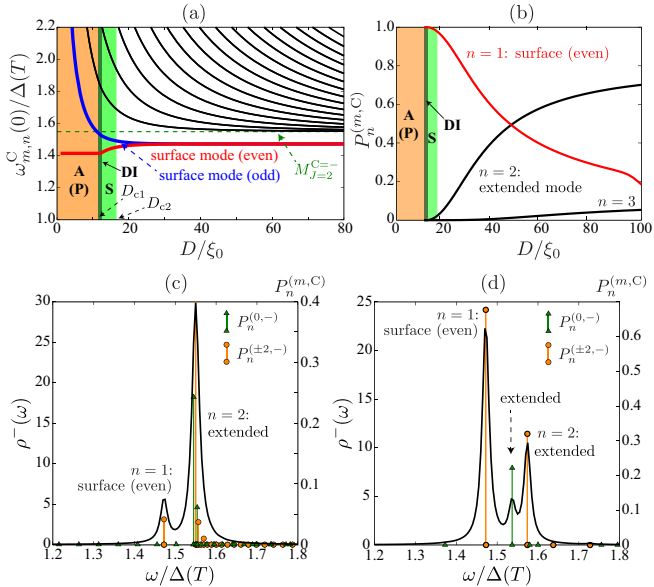}
\caption{
(a) Bosonic mode frequencies, $\omega^{\rm C}_{m,n}(0)$, as a function of 
    confinement, $D/\xi_0$. 
(b) Intensities, $P^{(m,{\rm C})}_n$, of the $n=1,2,3$ modes for $(m,{\rm C})=(\pm 2,-1)$ as a functions of $D/\xi _0$. ``S'' and ``A (P)'' stand for the regions in which the ``stripe'' and A (planar) phases are thermodynamically stable; ``DI'' indicates the region of dynamical instability with ${\rm Im}(\omega^{\rm C}_{m,n})>0$. 
	(c,d) Attenuation spectra, $\rho^-(\omega)$, and intensities of $m=0^+$ and $\pm 2$ modes at (c) $D/\xi_0=137$ (c) and (d) $D/\xi_0=31$, both at $T/T_{{\rm c}0}=0.8$ ($\gamma = 0.01\Delta (T)$).
}
\label{fig:soundm}
\end{figure}

To calculate the bound-state energy, consider a semi-infinite system with a specular surface ($z=0$) and approximate the local distorsion as $\epsilon\equiv f^2_{\parallel}-1\sim 0.2$ and $1-f^2_z(z)={\rm sech}^2(z/\lambda)$. For ${\rm C}=-$, Eq.~\eqref{eq:eom} has the bound state solution, 
$\omega^-_{\pm 2}(0)=\sqrt{M^2_{2,-}-(\frac{4}{5}-\frac{16}{5}\epsilon)\Delta^2(T)} < M_{2,-}$.
 The spatial profiles of the two lowest modes, obtained by numerically diagonalizing Eq.~\eqref{eq:eom}, are shown in the insets of Figs.~\ref{fig:surface}(a) and \ref{fig:surface}(b). 
The lowest mode for ${\rm C}=-$ is tightly bound to both surfaces on the scale of $5\xi_0$, while the ${\rm C}=+$ sector has the wave function extended throughout the system. We refer to the former (latter) as the surface bosonic mode (extended bosonic mode). In Fig.~\ref{fig:soundm}(a), we plot the thickness dependence, $D$, of the low-lying bosonic spectral weights. Note that irrespective of $D/\xi_0$ the surface bound modes always have a mass gap below that of the extended state. 

We also find that surface bosonic modes appear in the sectors of $(m,{\rm C})=(\pm 1,-)$, $(\pm 2,-)$, and $(0^+,+)$~\cite{miz18}. These modes are relatively insensitive to strong-coupling corrections to the GL parameters in $f_{\rm b}$, but are sensitive to the surface boundary condition. For retro-reflective boundaries, which induce maximal surface depairing such that $f_{\parallel}=f_z=\tanh(z/\lambda)$, the potential, $U^{\rm C}_m(z)$, becomes attractive, and thus bound-state solutions appear even for $(m,{\rm C})=(\pm 2,+)$. 

{\it Ultrasound attenuation.}--- Ultrasound spectroscopy has proven to be a powerful probe of the bosonic modes of bulk $^3$He \cite{hal90,mck90}.
Here we discuss the coupling, selection rules and signatures of the bosonic modes in confined $^3$He-B for ultrasound propagation and attenuation measurements. For this purpose, we introduce the effective coupling of order parameter fluctuations to the density fluctuation, $\rho(x)$,
\begin{align}
S_{\rm source} = 
- \int d^4x \rho(x) \hat{Q}_{\mu}\hat{Q}_i\left(
\beta^-\mathcal{D}^-_{\mu i}(x) 
+ 
\beta^+\mathcal{D}^+_{\mu i}(x)
\right).
\label{eq:source}
\end{align}
In Eq.~\eqref{eq:source}, $\hat{Q}_{\mu} \equiv -i\partial _{\mu}$ is the momentum operator and $\beta^{\rm C}$ 
represents the coupling strength between a density fluctuation and the order parameter. The term with 
$\beta^{+}$ 
was introduced in Ref. \cite{sal89b} to calculate the splitting of the $J^{\rm C}=2^+$ modes caused by the combined effects of magnetic field, dispersion, and texture. The action defined in Eq.~\eqref{eq:source} reproduces the coupling of bulk bosonic modes to ultrasound waves to leading order in $v_{\rm F}Q/\omega \ll 1$~\cite{miz18}.

The effective coupling in Eq.~\eqref{eq:source} leads to the addition of a source term in Eq.~\eqref{eq:H}, i.e. $[\omega^2-\mathcal{H}^{\rm C}_{\mu i,\nu j}]\mathcal{D}^{\rm C}_{\nu j} = \beta^{\rm C}(c_{\rm l}Q_{\parallel})^2b_{\mu i}$, where $b_{\mu i}=\hat{q}_{\mu}\hat{q}_i$ with $\hat{\bm q} = {\bm Q}/|{\bm Q}|$ being the sound propagation direction. 
The spectral density for the absorption of sound is defined as
$\rho(\omega)=-\frac{1}{\pi}\sum_{\rm C}\int dz\,
{\rm Im}\left[b^{{\rho}\ast}_{\mu i}\mathcal{D}^{\rm C}_{\mu i}(z,Q_{\parallel})\right]
\equiv
\rho^-(\omega) + \zeta^2 \rho^+(\omega)
$
 , where $\zeta\approx k_{\rm B}T_c/E_f\sim 10^{-3}$ 
 is the particle-hole asymmetry coupling that lifts the selection rule for coupling density fluctuations to the $C=+1$ bosonic modes \cite{koc81}.
For $v_{\rm F}Q_{\parallel}/\omega \ll 1$, the spectral functions reduce to 
\beq\label{eq:rho}
\rho^{\rm C}(\omega) = - \frac{1}{\pi}{\rm Im}
\sum _n \frac{|P^{(m,{\rm C})}_n|^2}{(\omega + i\gamma)^2-[\omega^{\rm C}_{m,n}(0)]^2}
\,,
\eeq
 where $\gamma\ll\omega^{\rm C}_{m,n}$ represents the natural linewidth of the mode. Equation~\eqref{eq:rho} has poles at the frequencies, $\omega^{\rm C}_{m,n}(0)$ corresponding to resonant absorption of sound by excitation of the bosonic mode $m,n,C$. In Eq.~\eqref{eq:rho}, $|P^{(m,{\rm C})}_n|^2$ represents the intensity of the coupling of the $n^{th}$ bosonic mode to the density.

Here we consider longitudinal sound propagating parallel to the surface ($\hat{\bm q}\parallel\hat{\bm x}$), where only the $m=0^+$ and $\pm 2$ modes contribute to $\rho(\omega)$.
In Fig.~\ref{fig:soundm}(b), we show results for $|P^{(\pm 2,{\rm -})}_n|$ as a function of $D/\xi _0$ at $T/T_{\rm c0}=0.8$. In the bulk limit, $D\gg100\xi_0$, the main contribution to $\rho^{-}(\omega)$ for $m=\pm 2$ originates from the transition to the $n=2$ mode that corresponds to the $J^{\rm C}=2^-$ modes in bulk $^3$He-B. As $D$ approaches the critical thickness $D_{\rm c2}$ at which the B to ``stripe'' transition occurs \cite{vor07}, the energy level of the $n=2$ extended mode becomes higher, while the surface bound mode with even parity remains the lowest mode (Fig.~\ref{fig:soundm}(a)). Figure \ref{fig:soundm}(b) shows that the surface bound mode dominates the main resonance in the sound attenuation spectrum for $D/\xi _0\lesssim 50$. 

Figures~\ref{fig:soundm}(c) and \ref{fig:soundm}(d) show the attenuation spectrum at $D/\xi _0 = 137$ and $31$, respectively. The crossover from the lowest extended mode to the surface bound mode gives rise to the shift of the main resonance peak in the attenuation spectrum, $\rho^-(\omega)$. Note that the resonance due to the surface bound mode exists as the satellite peak even for weak confinement, $D/\xi_0 = 137$, while the main peak shifts to the resonance of the surface bound mode at $D/\xi _0 =31$.

\begin{figure}[t!]
\includegraphics[width=85mm]{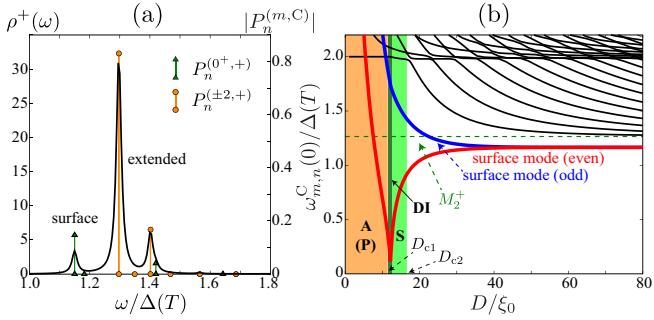}
\caption{(a) Ultrasound attenuation spectrum, $\rho^+(\omega)$, at $D/\xi _0 = 31$ ($\gamma = 0.01\Delta (T)$). (b) Bosonic excitation spectra for $(m,{\rm C})=(0^+,+)$ as a function of $D/\xi _0$. All calculations for $T/T_{{\rm c}0}=0.8$.}
\label{fig:soundp}
\end{figure}

Confinement also leads to a splitting of the absorption peak into two peaks for $\omega \approx M^-_2\sim 1.55 \Delta(T)$ and $D/\xi _0 \lesssim  31$. The origin of the splitting is this: the confinement potential explicitly breaks the $\point{SO(3)}{J}$ symmetry of the bulk $^3$He-B, which lifts the fivefold degeneracy of the $J=2$ modes and permits the hybridization of the $(J,m_J)=(2,0)$ mode with the $(0,0)$ phase mode in the $m=0^+$ sector. In contrast, an admixture of $m_J=\pm 2$ with other modes is prohibited by the remaining ${\tt SO(2)_{{J_z}}}$ symmetry. This leads to the mass splitting of the $m=0^+$ and $\pm 2$ extended modes, and is the  analogue of crystal-field splitting of energy levels of ions embedded in the solid state, where confinement plays the role of the crystal field. This effect may be observable through the splitting of sound attenuation peaks as seen in Fig.~\ref{fig:soundm}(d). 

The attenuation spectrum, $\rho^+(\omega)$, is displayed in Fig.~\ref{fig:soundp}(a). In the ${\rm C}=+$ sector, the surface bosonic mode exists only for $m=0^+$. The satellite peak with $\omega/\Delta \approx 1.15$ in Fig.~\ref{fig:soundp}(a) originates from resonant excitation of the surface mode. The pronounced resonance peak due to the $m=0^+$ extended mode is observed even in the thin slab limit with $D/\xi _0 = 31$. As shown in Fig.~\ref{fig:soundp}(b), the spectrum of the surface mode with even parity exhibits anomalous behavior with decreasing $D/\xi _0$; the mass gap closes for $D\sim D_{\rm c1}$, and the mode amplitude smoothly evolves from a surface localized mode to an extended mode as $D$ approaches $D_{\rm c1}$. We demonstrate below that the anomalous behavior of the $m=0^+$ mode is a signature of the dynamical instability of the confined ground state in Eq.~\eqref{eq:bw} towards a spatially modulated superfluid phase, i.e. the stripe phase~\cite{vor07}.

\begin{figure}[t!]
\includegraphics[width=85mm]{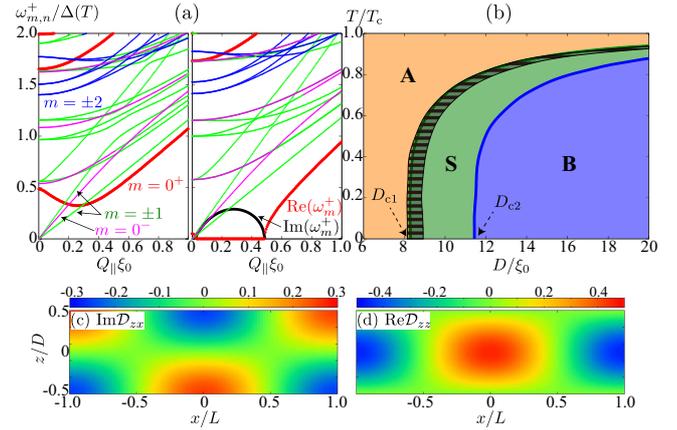}
\caption{(a) Mode dispersions at $D/\xi _0=9.4$ (left) and $8.8$ (right), for $T/T_{{\rm c}0}=0.4$. Eigenvalues for the $0^{+}$ mode become pure imaginary for $D/\xi =8.8$, where ${\rm Im}\omega$ is denoted by the thick (black) curve. (b) $T$-$D$ phase diagram in the weak coupling limit. The stripe-shaded area corresponds to the dynamical instability region of the translationally invariant B-phase. 
The data on the phase diagram are taken from Ref.~\cite{wim16}. 
(c,d) Spatial profiles of unstable modes: 
(c) ${\rm Im}\mathcal{D}_{zx}(z)\cos(Q_{\parallel}x)$ and 
(d) ${\rm Re}\mathcal{D}_{zz}(z)\cos(Q_{\parallel}x)$.}
\label{fig:stripe}
\end{figure}

{\it Dynamical instability of the surface bound mode.}--- In Fig.~\ref{fig:stripe}(a), we show the dispersion of the low-lying bosonic modes for confinement near $D_{\rm c1}$ at $T=0.4T_{\rm c0}$. For the ground state we assume translational symmetry parallel to the surface as in Eq.~\eqref{eq:bw}. The mass gap of the $m=0^+$ branch closes near $D_{\rm c1}$ and its dispersion softens as $D$ decreases. At the point where the branch dispersing from the $m=0^+$ massive mode touches zero frequency a pair of conjugate pure imaginary frequencies appear in the region of small momenta, $0 \lesssim Q_{\parallel}\xi _0\lesssim 0.5$, at $D= 8.8 \xi _0$. As $D$ further decreases, the momentum region with ${\rm Im}\omega\neq 0$ gradually narrows. For $D\lesssim 8.3 \xi _0$ the imaginary-frequency modes vanish and all bosonic modes are gapped out again.

The eigenvalues of Eq.~\eqref{eq:eom} reflect the curvature of $f_{\rm GL}$ around the translationally invariant equilibrium state of Eq.~\eqref{eq:bw}. 
%
%
The appearance of pure imaginary eigenvalues at finite $Q_{\parallel}$ is the signature of negative curvature of $f_{\rm GL}$ and the exponential growth in time of the spatially modulated bosonic $0^{+}$ mode; $\mathcal{D}^{\rm C}_{\mu i}(x) = \sum _{{\bm Q}_{\parallel}\in\mathbb{Q}} e^{\pm |{\rm Im}\omega({\bm Q}_{\parallel})| t +i{\bm Q}_{\parallel}\cdot{\bm r}} \mathcal{D}^{\rm C}_{\mu i}({\bm Q}_{\parallel},z)$, where $\mathbb{Q}:=\{ {\bm Q}_{\parallel}| {\rm Im}\omega({\bm Q}_{\parallel})\neq 0 \}$. The softening at finite ${ Q}_{\parallel}$ indicates that Eq.~\eqref{eq:bw} is dynamically unstable toward a ``crystallization'' of the superfluid vacuum with the spatial period $L\equiv \pi/Q_{\parallel}$. %
In Figs.~\ref{fig:stripe}(c) and \ref{fig:stripe}(d) we plot the dominant components of the unstable modes with characteristic wavenumbers $\pm Q_{\parallel}$, i.e. ${\rm Im}\mathcal{D}_{zx}(x,z)$ and ${\rm Re}\mathcal{D}_{zz}(x,z)$. The other components are negligible.
The spatial modulation of the unstable mode coincides with that of the stripe phase~\cite{vor07,wim16}.

Figure~\ref{fig:stripe}(b) shows the dynamically unstable region projected onto the $D$-$T$ phase diagram in the weak-coupling limit. The unstable eigenmodes appear in the stripe-shaded area, which covers the phase boundary between A and stripe phases for all temperatures. As shown in Ref.~\cite{wim16}, the stripe phase is sensitive to strong-coupling corrections to $f_{\rm b}$. In a separate analysis we show that the dynamical instability region is less sensitive to strong coupling corrections, and stays in the vicinity of the phase boundary between the B and planar phases~\cite{miz18}. Thus, acoustic excitation of the planar-distorted B phase, tuned toward the region of the dynamical instability, may provide a mechanism for stabilizing the crystalline stripe phase.

{\it Summary and Outlook.}--- Using an effective field theory we reveal a rich spectrum of bosonic excitations in confined $^3$He-B, including surface bound bosonic modes, splitting of the fivefold $J=2$ modes, and the dynamical instability signalling the formation of a crystalline stripe phase. All of these effects of confinement on the bosonic mode spectrum can be captured through ultrasound attenuation spectroscopy. 

The bosonic effective field theory defined by $S_{\rm B}$ does not incorporate the spectrum of helical MFs. The full action for $^3$He-B, which is a prototype of DIII topological SCs, is given by $S_{\rm B}+S_{\rm F}+S_{\rm FB}$, where $S_{\rm F}$ and $S_{\rm FB}$ are the actions for the spin-1/2 Majorana field, $\chi$, and the fermion-boson coupling, respectively. The surface MF in $^3$He-B obeys $P_3\chi=-\chi$, where $P_3$ is the combined symmetry of the time reversal and $C_2$ rotation about $\hat{\bm z}$~\cite{miz12a,miz16}. The $P_3$ symmetry protects the characteristic helical spin-orbit texture of the MFs, and as we show in a separate report~\cite{miz18}, the fermion-boson coupling described by $S_{\rm FB}$ is subject to a strict selection rule. In the ${\rm C}=-$ sector, only the $m=\pm 1$ bosonic modes couple to the helical MFs: the decay of the $m=\pm 2$ surface bosonic modes into helical MFs is forbidden by the $P_3$ symmetry.

Lastly, we note that surface bosonic modes are not unique to $^3$He-B but are expected to be common to SCs with multi-component order parameters as they often support long-lived midgap bosonic modes, and an attractive confinement potential, as in Eq.~\eqref{eq:eom}, generated by the local distorsion of the equilibrium order parameter. As an example consider the two-dimensional representations for cubic and hexagonal crystal symmetries, whose bulk GL functional is given by
$f_{\rm b}=-\frac{N_{\rm F}}{3}(1-\frac{T}{T_{\rm c0}})(|\eta _1|^2+|\eta _2|^2)
+\beta_1 (|\eta_1|^2+|\eta_2|^2)^2+\beta_2 (\eta^{\ast}_1\eta_2-\eta_1 \eta^{\ast}_2)^2$~\cite{sig91}.
The equilibrium state for $\beta_2>0$ is $(\eta_1, \eta_2)=(\eta _0,0)$. In the bulk, the masses of the bosonic modes, $\mathcal{D}_1\equiv \eta _1-\eta _0$ and $\mathcal{D}_2\equiv \eta_2$, are given by $(M^{\rm C}_j)^2\equiv \frac{1}{\tau}\frac{\delta^2f_{\rm b}}{\delta \mathcal{D}^{\rm C}_j \delta \mathcal{D}^{\rm C}_j}$. Thus, there exist two NG bosons with $M^{-}_1=M^{+}_2=0$ and two massive bosons with $M^-_2/M^+_1=\sqrt{\beta_2/2\beta_1}$. The local distortion of the equilibrium order parameter near the surface, $\eta_0(z) < \eta_0$, generates the attractive potential for $\mathcal{D}^-_2$ as $\frac{1}{\tau}\frac{\delta^2f_{\rm b}}{\delta \mathcal{D}^-_2 \delta \mathcal{D}^-_2}=(M^-_2)^2-\frac{2}{\tau}\beta_2(\eta^2_0-\eta^2_0(z))$. Hence, the surface bound bosonic modes emerge on the surface of these classes of unconventional SCs, and are expected to be observable through their coupling to transverse electromagnetic waves~\cite{hir89,yip92,sau15}.

We thank J. J. Wiman for sharing numerical results of the stripe phase diagram. 
The work of T.M. was supported by Japan Society for the Promotion of Science (JSPS) (Grant No.~JP16K05448) and ``Topological Materials Science'' (Grants No.~JP15H05855 and No.~JP15K21717) KAKENHI on innovation areas from JSPS of Japan. 
The research of J.A.S. was supported by the National Science Foundation (Grants DMR-1106315 and DMR-1508730).
 This work was initiated at the Aspen Center for Physics, which is supported by National Science Foundation grant PHY-1607611.


%
\end{document}